\documentclass[runningheads]{llncs}
\usepackage[T1]{fontenc}

\usepackage[english]{babel}
\usepackage{blindtext}
\usepackage{orcidlink}
\usepackage{amssymb}
\usepackage{tabularx}
\usepackage{booktabs}
\usepackage[normalem]{ulem}
\usepackage{pgfplots}
\pgfplotsset{compat=1.18}

\newcommand{\BG}{{BGProtect}}
\newcommand{\ignore}[1]{}

\title{AI Censorship title}

\usepackage{mathtools} 
\usepackage{bbm} 
\usepackage{booktabs} 
\usepackage{tabularx} 
\usepackage{makecell} 

\usepackage{enumitem}
\usepackage{url}
\usepackage{subcaption}
\usepackage{pgf-pie}

\begin{document}

\title{Measuring DNS Censorship of Generative AI Platforms}

\author{
    Harel Berger\inst{1}\orcidID{0000-0001-6035-5127} \and
    Yuval Shavitt\inst{2}\orcidID{0000-0002-0701-2405}
}

\authorrunning{H. Berger and Y. Shavitt}

\institute{
    Department of Computer Science, Georgetown University, Washington DC, USA \\
    \email{hb711@georgetown.edu} \and
    School of Electrical Engineering, Tel Aviv University, Tel Aviv, Israel \\
    \email{shavitt@eng.tau.ac.il}
}

\maketitle

\begin{abstract}
Generative AI is an invaluable tool, however, in some parts of the world, this technology is censored due to political or societal issues. In this work, we monitor Generative AI censorship through the DNS protocol. We find China to be a leading country of Generative AI censorship. Interestingly, China does not censor all AI domain names. We also report censorship in Russia and  find inconsistencies in their process.
We  compare our results to other measurement platforms (OONI, Censored Planet, GFWatch), and present their lack of data on Generative AI domains.
\end{abstract}

\section{Introduction}
In recent years, Generative AI emerged as a great tool for millions of users around the world~\cite{gpt_news}. 
The pursuit of allowing equal use of AI (termed "AI for all"~\cite{avellan2020ai,ramos2021ai}) seeks an equal chance for every world citizen to enjoy the benefits of this technology.  However, some countries block access to Internet content and services via Internet Censorship~\cite{warf2011geographies,aceto2015internet,zittrain2017shifting}. Internet Censorship follows different methods, e.g., block pages~\cite{zittrain2017shifting,ververis2021understanding,jones2014automated}, forged DNS responses~\cite{hoang2021great,tang2016depth,farnan2016poisoning}, and various connection errors~\cite{zittrain2017shifting,master2023worldwide}. Most of the censorship methods try to identify clear-text network communication packets. However, sophisticated censorship methods tackle the identification of encrypted data~\cite{wu2023great}.

Several countries have been identified as heavily censoring the Internet for their citizens, such as China~\cite{deibert2008access,griffiths2021great,clayton2006ignoring,wu2023great,bambauer2005internet}, Russia~\cite{ermoshina2022market,duffy2015internet}, North Korea~\cite{kreishan2022media,chen2010north}, Iran~\cite{hashemzadegan2022internet,aryan2013internet,bagheri2023press} and  Turkmenistan~\cite{nourin2023measuring,bohr2008turkmenistan}. One of the most highly-invested and deeply explored state-censorship systems is "The Great Firewall of China" (GFW) \cite{clayton2006ignoring,hoang2021great,tang2016depth,griffiths2021great}, dated back to the early 2000's~\cite{walton2001china}. 
In the context of AI tools even some democracies, such as Italy\footnote{https://www.cnbc.com/2023/04/04/italy-has-banned-chatgpt-heres-what-other-countries-are-doing.html}, vowed to block AI tools. The European Union and the United States also made the first steps in regulating these tools~\cite{cheng2024unravelling}.

In this work, we aim to discover the level of censorship and blocking of various Generative AI tools. For this aim, we build lists of leading tools in several categories such as chatbots, image generators, text-to-sound, text-to-video, AI playgrounds, and more.  For each website of a tool in the lists, we examine the level of DNS blocking by sending DNS queries to more than 850,000 DNS resolvers, covering almost every nation on Earth, and identify the resolvers that do not respond with the correct IP address.

Our results show that China is leading the world in AI censorship, practically blocking sites such as Huggingface, ChatGPT, Poe, and Llama. Blocking is done, like reported in the past for other Chinese censorship \cite{lowe2007great,stadler2023survey}, by returning IP addresses selected from a list of global ISPs such as Facebook, Twitter, Dropbox, and more.

In Russia some domains are blocked, but only by some leading ISPs. TransTelecom, Cosmos, ER-Telelcom, and RUNNET (The Russian university network), are blocking the DNS queries of Llama and CourseHero, while others, such as VimpelCom (Beeline) and Rostelecom, are not.
Blocking is done by returning an IP of a blocking site, a nonexisting domain, or an internal IP address.
Interestingly, blocking of Llama in Russia is not always done for the Llama domain name, llama.meta.com, but for the first cname used in the LLama resolution chain, e.g., star.c10r.facebook.com. Sometimes, the blocking is done on a cname of a cname (fz139.ttk.ru). 

The rest of the paper is as follows; First, Section~\ref{related} discusses related work. Second, Section~\ref{methods} describes data collection. Next, Section~\ref{results} presents our results. Then, 
Section~\ref{others} compares our findings with data from previous tools.
Section~\ref{conclusion} concludes our work and paves the way for future work.

\section{Related Work}
\label{related}
A large volume of work has already delved into the nature of Internet censorship~\cite{warf2011geographies,aceto2015internet,zittrain2017shifting,ververis2021understanding,jones2014automated,wu2023great,deibert2008access,griffiths2021great,clayton2006ignoring,wu2023great,bambauer2005internet,ermoshina2022market,duffy2015internet,kreishan2022media,chen2010north,nourin2023measuring,bohr2008turkmenistan}. Prior work examined several forms of DNS-based Internet censorship, ranging from the use of block page~\cite{zittrain2017shifting,ververis2021understanding,jones2014automated}, using fake DNS response~\cite{hoang2021great,tang2016depth,farnan2016poisoning} and connections errors~\cite{zittrain2017shifting,master2023worldwide}. Several countries stand out as significant censors, such as China~\cite{deibert2008access,griffiths2021great,clayton2006ignoring,wu2023great,bambauer2005internet}, Russia~\cite{ermoshina2022market,duffy2015internet}, North Korea~\cite{kreishan2022media,chen2010north}, Iran~\cite{hashemzadegan2022internet,aryan2013internet} and Turkmenistan~\cite{nourin2023measuring,bohr2008turkmenistan}. China is flagged as the most explored country in this notion, as its censorship methodology won the term "The Great Firewall of China" (GFW)~\cite{clayton2006ignoring,hoang2021great,tang2016depth,griffiths2021great,wu2023great}.

Several works analyzed the nature of GFW. Deibert et al.~\cite{deibert2002dark} analyzed the blocking policy of the GFW and its centrality. Using standard A-type DNS queries, GFWatch~\cite{hoang2021great} analyzed the inner regular expressions used by GFW and GFW's impact on the DNS ecosystem. Wu et al.~\cite{wu2023great} analyzed the ability of GFW to censor encrypted data, ruling out non-encrypted data based on efficient heuristics.
The recent GFWeb~\cite{298156} leveled up GFW's measurement, allowing the measurement of millions of domains for a long period of time.
These previous works focused on the large-scale view of GFW's censorship. We aim to analyze specifically the topic of Generative AI censorship, a recent impactful topic. We also differentiate between different Generative AIs in GFW's censorship. 

Censorship measurement platforms have been established in the past. ICLab \cite{ICLab:SP20}, Censored Planet~\cite{sundara2020censored} and OONI~\cite{filasto2012ooni} explored censorship globally using different protocols, including DNS, TCP, HTTP, and HTTPS. Other platforms such as GFWatch~\cite{hoang2021great} and GFWeb~\cite{298156} focus on China's GFW censorship. We compare our results to the findings of these tools, when available since these tools do not focus on Generative AI domains. 

A recent work explored the challenges of Generative AI in China~\cite{cheng2024unravelling}. However, this work only raised the challenges of China's censorship of Generative AIs and did not practically evaluate them. Also, it did not evaluate other countries' practices of Generative AI censorship, as we do in this work. 

In summary, previous studies did not focus on the censorship of Generative AI in the world. We explore these domains, as their number is growing daily and so is their utility. 

\section{Methods}\label{methods}
In this section, we present the data collection. Then, we describe how we picked the Generative AI domains. The ethics considerations are disclosed in Appendix~\ref{ethics}.
\subsection{Data Collection}
\label{data-col}
We explore Internet censorship through the lens of the DNS protocol, a common practice in this topic~\cite{pearce2017global,aryan2013internet,298156,hoang2021great}.
To perform the DNS queries, we use a tool from \BG~\cite{bgprotect}, which uses proxies around the world to enable large-scale query campaigns.  
The vast majority of queries are done through proxies that reside in three cloud providers with locations around the world: The Netherlands (Amsterdam), Great Britain (London), Poland (Warsaw), Canada (Toronto), and the USA (Chicago, Los Angeles, Dallas, Phoenix, Miami, Atlanta, Ashburn, and NYC).  Only $\sim0.5\%$ of the queries were done via residential proxies, however, none were from China or Russia.

In this study, we only use A queries.
The system queries over 850,000 DNS recursive resolvers that were found to be mostly reliable, namely they almost always answer queries to known domain names, correctly.

The DNS resolvers are queried at rates that increase according to the resolver's importance according to a proprietary algorithm, such that Google's 8.8.8.8 service queries every 20 minutes while the resolver of a small local AS is queried at a rate 250 slower. The rate of queries is not significant since almost all of the resolvers show consistent behavior, namely they either block a DNS or return the correct IP address (as exampled in Table~\ref{tab:correct_china} and Table~\ref{tab:fake_china}). We queried the domains for $\sim$2 months, between March 1, 2024, and April 29, 2024.

The geolocation of the resolvers was obtained using the \BG~database. 
In general, DNS resolver location is quite easy to get at the national level. Note that we care about the geolocation of resolvers from nations where censorship occurs to ensure our results are correct. For example, an error in the geolocation of a resolver between Canada and France is insignificant for our study since these countries are not censoring any of the domain names we checked.

For China and Russia, the countries where most Generative AI censorship occurs, and for other resolvers that returned a wrong answer to our queries, we also performed random manual geolocation verification. Our sample biased resolvers based on their importance as appears by the query rate of the \BG~system.  There were only a few cases where we found location errors.  The most common errors are:
\begin{itemize}
    \item Distinguishing between China and Hong Kong (e.g., Guangzhou and Shenzhen are only a few milliseconds away from Hong Kong).
    \item Geolocation of an anycast IP address in one location.
\end{itemize}
\subsection{Representative Domains}

\begin{table*}[htbp]
\small
\centering
\begin{tabular}{|l|l|l|l|l|}
\hline
\textbf{Chatbots}     & \textbf{T2I }        & \textbf{T2S }                    & \textbf{Assistant} & \textbf{Work}         \\ \hline
chat.openai.com       & midjourney.com                       & murf.ai                                          & app.grammarly.com                   & deepai.tn                              \\ \hline
claude.ai             & openai.com                           & lovo.ai                                          & wordtune.com                        & jasper.ai                              \\ \hline
bard.google.com       & dreamstudio.ai                       & resemble.ai                                      & coursehero.com                      & bardeen.ai                             \\ \hline
copilot.microsoft.com & civitai.com                          & soundraw.io                                      & designs.ai                          & simplified.com                         \\ \hline
llama.meta.com        & starryai.com                         & \textbf{T2V}                    & brainly.com    & copy.ai                                \\ \hline
janitori.com          & openart.ai                           & synthesia.io                                     & originality.ai & \textbf{Multimodal}   \\ \hline
beta.character.ai     & \textbf{Playground} & colossyan.com                                    & quillbot.com                        & writesonic.com                         \\ \hline
kimi.moonshoot.cn     & poe.com                              & hourone.ai                                       & rephrase.ai                         & deepai.org                             \\ \hline
simsimi.com           & sdk.vercel.ai                        & \textbf{LLM generation} &       & runwayml.com                           \\ \hline
replika.com           & chat.lmsys.org                       & rasa.community                                   &                            & \textbf{Storytelling} \\ \hline
01.ai                 & \textbf{Community}  & cohere.com                                       &                            & novelai.net                            \\ \hline
deeppavlov.ai         & huggingface.co                       & scale.com                                        &                           &                \textbf{S2T}                        \\ \hline
grok.x.ai             & \textbf{I2T}        & zerogpt.com                                      &                       &             serve.ivrit.ai                           \\ \hline
mistral.ai            & ultralytics.com                      & copy.ai                                          &                                     &                                        \\ \hline
\end{tabular}
\caption{\textbf{Generative AI domains used in this study by categories} T2I refers to Text-to-Image. I2T reflects Image-to-text. T2S is an acronym for Text-to-sound. T2V refers to test-to-voice. S2T reflects sound-to-text.}
\label{tab:tab-cat1}
\end{table*}

Our method for identifying censorship is via DNS resolution.
Thus, we can not test the blocking of inner paths within a domain name.  This limits us from testing the blocking of projects in git or the DALLE/SORA models of openAI. For the latter domains, we tested their parent domain - OpenAI.   
We enumerated a sample of 61 Generative AI domains we explored in our research in Table~\ref{tab:tab-cat1}. Some of the categories are more popular in the world (e.g., chatbots, AI assistants, and text-to-image (T2I)), therefore we sampled more domains in these categories. We also sampled some AIs associated with countries to see if a specific country is targeted, for example, Kimi in China, Deepai in Tunisia, and Ivrit in Israel.

\section{Results} \label{results}

In this section, we describe the findings of Generative AI blocking. These findings are based on a $\sim$two-month probe, from March 1, 2024, to April 29, 2024.
The nations with significant blocking at the DNS level are China and Russia. Although it was found that China and Russia benefited from AI in their digital warfare against the USA~\footnote{https://www.nbcnews.com/politics/national-security/china-russia-ai-divide-us-society-undermine-us-elections-power-rcna142880}, both countries are less positive towards their citizens using these technologies. This section refers to a blocking event as a consistent blocking behavior over time. We disregard short-time events, where events occurred for less than a day. Also, the last octet of an IP address may change based on load balancing and inner changes inside a domain. Thus, we map IPs to their respective ASNs and compare the correct IP-$>$ASN map\footnote{More details about the IP to ASN mapping, see \cite[Sec.~III]{SASA}.} to the correlative ASNs of the results we obtained.

\subsection{Results Consistency}
Since in the data we got, the resolvers are sampled at different rates, we show in Tables~\ref{tab:correct_china} and \ref{tab:fake_china} that this does not affect the final outcome.  The tables show that the percentage of good answers for censored domain names and bad answers for non-censored domains is not changed by the sampling method: either examining all the results obtained over a week or taking only the last result per resolver.

The tables show 5 censored domains and 3 non-censored domains, where we show results only from Chinese resolvers.  
For censored domains (Table~\ref{tab:correct_china}), the correct answers are limited to less than 1\%. Note that for Llama we got 3\% "correct" answers since China's censorship is done using top-known ASNs including Llama correct ASN (AS32934)~\cite{hoang2021great,tang2016depth,farnan2016poisoning}. Our analysis is done at the ASN level hence the 2\% increase in "correct" answers.
For the uncensored domains (Table~\ref{tab:fake_china}), the bad responses are less than 0.1\% for both the last probe and the sum of all probes.
Thus, we can conclude that the results are independent of the resolver sampling method.

\begin{table}[h]
\begin{tabularx}{\columnwidth}{|c|*{4}{>{\centering\arraybackslash}X|}}
\hline
\textbf{Domain} & \multicolumn{2}{c|}{\textbf{Last Response}} & \multicolumn{2}{c|}{\textbf{All Responses}} \\
\cline{2-5}
& \textbf{Good} & \textbf{All} & \textbf{Good} & \textbf{All} \\
\hline
huggingface.co & 903 (0.72\%) & 124,728 & 2,092 (0.82\%) & 255,892 \\
\hline
dreamstudio.ai & 357 (0.29\%) & 124,728 & 996 (0.39\%) & 255,868 \\
\hline
civitai.com & 1,083 (0.87\%) & 124,728 & 2,292 (0.9\%)& 255,694 \\
\hline
llama.meta.com & 182 (0.14\%) & 124,728 & 11,563 (3.09\%) & 296,888 \\
\hline
chat.openai.com & 1,034 (0.83\%) & 124,728 & 2,708 (1.06\%) & 256,234 \\
\hline
\end{tabularx}
\caption{Constistency of correct answers for censored domains in China. The timeframe of these probes is between March 24, 2024, to March 31, 2024. 
We present the results of the correct responses (Good) and all responses (All), in the view of the last probe (Last) and the sum of all probes together (All).}
\label{tab:correct_china}
\end{table}

\begin{table}[h]
\begin{tabularx}{\columnwidth}{|c|*{4}{>{\centering\arraybackslash}X|}}
\hline
\textbf{Domain} & \multicolumn{2}{c|}{\textbf{Last Response}} & \multicolumn{2}{c|}{\textbf{All Responses}} \\
\cline{2-5}
& \textbf{Bad} & \textbf{All} & \textbf{Bad} & \textbf{All} \\
\hline
bard.google.com & 2 (0.002\%) & 124,728 & 4 (0.001\%) & 258,072 \\
\hline
kimi.moonshoot.cn & 7 (0.006\%) & 124,728 & 19 (0.006\%) & 301,929 \\
\hline
www.01.ai & 5 (0.004\%) & 124,728 & 12 (0.004\%) & 304,718 \\
\hline
\end{tabularx}
\caption{Constistency of bad answers for non-censored domains in China. The timeframe of these probes is between March 24, 2024, and March 31, 2024. We present the results of the bad response (Bad) and all responses (All) per the last probe (Last) and sum all probes together (All).}
    \label{tab:fake_china}
\end{table}

\subsection{Blocking in China }
\label{china-block}
\begin{table}[]
\centering
\begin{tabular}{|c|c|c|l|}
\hline
\textbf{Domains/Responses} & \textbf{Correct} & \textbf{Incorrect }&\textbf{Errors}
\\ \hline
\textbf{llama.meta.com}    & 0.04    & 0.12      & 0.84                 \\ \hline
\textbf{chat.openai.com}   & 0.01    & 0.97      & 0.02                 \\ \hline
\textbf{chat.lmsys.org}    & 0.01    & 0.97      & 0.02                 \\ \hline
\textbf{beta.character.ai} & 0.01    & 0.97      & 0.02                 \\ \hline
\textbf{poe.com}           & 0.01    & 0.97      & 0.02                 \\ \hline
\textbf{huggingface.co}    & 0.01    & 0.98      & 0.01                 \\ \hline
\textbf{dreamstudio.ai}    & 0.01    & 0.19      & 0.8                  \\ \hline
\textbf{civitai.com}       & 0.01    & 0.97      & 0.02                 \\ \hline
\textbf{starryai.com}      & 0.01    & 0.97      & 0.02                 \\ \hline
\textbf{openart.ai}        & 0.01    & 0.19      & 0.8                  \\ \hline
\textbf{novelai.net}       & 0.02    & 0.18      & 0.8                  \\ \hline
\textbf{coursehero.com}    & 0.01    & 0.98      & 0.01                 \\ \hline
\textbf{writesonic.com}    & 0.01    & 0.21      & 0.78                 \\ \hline
\textbf{deepai.org}        & 0.01    & 0.98      & 0.01                 \\ \hline
\end{tabular}
\caption{Blocking Statistics of Generative AIs by China. The percentages shown below are from 124,728 probed resolvers in China. We tabulate the portion of correct answer, incorrect answer, and errors (timeouts, non-exist domains, etc,). The described behaviors were found to be consistent over time.}
\label{tab:china_blocks}
\end{table}

In China, we queried 124,728 resolvers for each of the 61 domains in Table~\ref{tab:tab-cat1}. We found 14 domains that we suspect are blocked by China. These results are based on an eight-week probe, from March 1, 2024, to April 22, 2024. We summarized the results in Table~\ref{tab:china_blocks}. We found that 9 out of 14 domains are mostly blocked (e.g., ChatGPT and Huggingface): 97-98\% of our queries come back with wrong answers, and only 1\% provided an IP address in the correct ASN. 1-2\% of the queries to these 10 domains return a failure or timeout. The other 5 domains (e.g., Llama and Dreamstudio) also show a low percentage of correct answers but show a large percentage (78-84\%) of timeouts, 12-21\% of the replies return a wrong IP address. Investigation with  \BG~shows that these timeout events are a result of their query system configuration, so most likely these 5 domains are blocked, as well. Unfortunately, we could not get additional data from \BG.

As previous studies of Chinese censorship found~\cite{hoang2021great,tang2016depth,farnan2016poisoning}, China's domain blocking is done by answering DNS queries with incorrect IP addresses that are announced by western ASNs; the top ones used are Facebook (AS32934), Twitter (AS13414), and Dropbox (AS19679). As a result the percentage of correct answers for llama.meta.com is higher (4\%) since our analysis is done by examining the ASN of the obtained IP address, disregarding the IP address itself. 

Interestingly, in Hong Kong and Macao, which are special administrative regions of China, none of the domains we checked are blocked.
For example, China blocks ChatGPT, with 97\% bad answers (Table~\ref{tab:china_blocks}). However, in Macao and Hong Kong, we found that only 0.3\%  and 0.2\%, respectively, of the resolvers block this domain. Most resolvers do not send incorrect responses in these areas, unlike mainland China. 
The clear difference though the proximity of Hong Kong and Macao to mainland China is an attestation of the quality of the geolocation we used, since Macao and to a greater extent Hong Kong are only a few milliseconds apart from the important Chinese hubs in Guangdong Province, most notably Guangzhou and Shenzhen.  

\subsubsection{ China: Distinction Between Related Domains}
\label{china-no-block}
China is found to be the most active censor of Generative AI.  Here we attempt to shed light on the distinction between related domains.
\begin{enumerate}
    \item \textbf{Chatbots} 
    \begin{enumerate}
        \item China blocks one specific Chatbot - ChatGPT.  OpenAI achievements on Generative AI make it hard for China AI dreams to become a reality\footnote{https://www.scmp.com/tech/big-tech/article/3253034/openais-sora-pours-cold-water-chinas-ai-dreams-text-video-advancements-prompt-more-soul-searching}. Also, China attempts to block ChatGPT is reasoned by China's claim of spreading of political propaganda by this Generative AI\footnote{https://www.theguardian.com/technology/2023/feb/23/china-chatgpt-clamp-down-propaganda}. This can be a reason for China to censor OpenAI and its products. However, China is not blocking the parent domain of ChatGPT: OpenAI, and not blocking the other OpenAI domains DALLE and  SORA. 
    \item Also, China does not censor all Western chatbots equally. China blocks a niche chatbot framework as character.ai. However, famous rivals of ChatGPT, such as Claude and Bard/Gemini are not censored at all in China. The only real rival of ChatGPT that China blocks is Facebook's Llama. 
    \end{enumerate}
    
\item \textbf{Playground} Out of the three AI playgrounds, China censors Poe.com and chat.lmsys. It does not censor Vercel. As these platforms help to compare and differentiate different Generative AI models, it can be natural that China will block users from accessing them. Most simply, these platforms grant access to ChatGPT (or an anonymous version of it, in the case of Chat.lmsys), which China blocks. However, Vercel, being one of these playgrounds is not blocked by China.
\item \textbf{T2I} China heavily censors this category. Dreamstudio.ai, Civatai.com, Starryai.com, and Openart.ai are all suspected as blocked/blocked by China. However, not all domains in this category are censored. Probably the most associated domain/tool of T2I, Midjourney, is not blocked by China. Also, as mentioned earlier, the host of DALLE, Openai.com, is not blocked. 
\item \textbf{AI assistants} Coursehero, a domain that hosts homework aid by AI is blocked by China. However, other assistant AIs such as Grammarly, Wordune, Quillbot, and others are not blocked by China.
\item \textbf{Multimodal} A growing body of AI models tries to generate content for a wide range of tasks, such as text, T2I, etc. These multimodal models are not treated the same by China. Writesonic.com and Deepai.org are suspected of being blocked/blocked by China. However, Runwayml.com is not blocked by it.
\end{enumerate}
In summary, the Chinese censorship of Generative AI differentiates between AI tools in the same category. This raises some question marks as to why a specific AI tool is blocked and similar tools are not.

\subsection{Blocking in Russia }
\label{russia-blocks-sec}

In Russia, we probed more than 59,000 DNS resolvers for two weeks, from April 15, 2024, to April 29, 2024. While we had not found nation-level blocking of any domain name, some domains were blocked by significant ISPs. Most notably, we found continuous blocking in  TransTeleCom (AS20485), one of the largest Russian ISPs, of two tested domains:\\ llama.meta.com and coursehero.com. Although Llama, as a part of Meta's website is expected to be blocked\footnote{https://www.theguardian.com/world/2022/mar/21/russia-bans-facebook-and-instagram-under-extremism-law}, we believe that the following analysis will shed light on the nature of the censorship of this domain, and Coursehero.
The same domain names are also continuously blocked by the Russian university network, RUNNet (AS3267).  In other smaller networks, we found similar blocking behavior of at least one of these two domain names.
On the other hand, many top Russian companies provide the correct resolution for these domain names, e.g., Rostelecom (AS12389), VEON (VimpelCom AS3216), and Megafone (AS31133).



Table~\ref{tab:russia_20485_3267_blocks} summarizes behavioral data for the blocking ASNs in Russia for llama.meta.com and coursehero.com. 
For each ASN, we tabulate the consistency of blockage: no block (\checkmark), continuously blocked ($\times$), and an alternation between blocked and correct answer ($\sim$). The right column (R) lists the number of resolvers queried for each ASN.

\begin{table}[h]
\centering
\begin{tabular}{|c|c|c|c|c|c|c|c|}
\hline
\textbf{Domains$\Rightarrow$} & \multicolumn{3}{c|}{\textbf{llama.meta.com}} & \multicolumn{3}{c|}{\textbf{coursehero.com}} & \textbf{R} \\
\hline
 \textbf{ASN$\Downarrow$}& \textbf{\checkmark} & \textbf{$\times$} & \textbf{$\sim$} & \textbf{\checkmark} & \textbf{$\times$} & \textbf{$\sim$} &\\
\hline
\textbf{AS3267}   & 0.7  & 0.2 & 0.1 & 0 & 1.0 & 0&10 \\
\hline
\textbf{AS20485}   & 0.415 & 0.4 & 0.185 & 0.73 & 0.19 & 0.08&140 \\
\hline
\textbf{AS8641}   & 0.875	&0.125	&0
 & 0.78 & 0.19 & 0.03 & 32\\
\hline
\textbf{AS41446}   & 0	&1.0	&0
 & 1.0 & 0 & 0 &1\\
\hline
\textbf{AS16047}   & 0	&1.0	&0
 & 0 & 1.0 & 0 &1\\
\hline
\textbf{AS33908
}   & 0	&0	&1.0
 & 0.5 & 0 & 0.5 &2\\
\hline
\textbf{AS43478
}   & 0.86	&0.14	&0
 & 0 & 1.0 & 0 &14\\
\hline
\textbf{AS199945
}   & 0.5	&0.5	&0
 & 0.25 & 0.5 & 0.25 &4\\
\hline
\textbf{AS39503
}   & 0.1	&0.8	&0.1
 & 0 & 1.0 & 0 &10\\
\hline
\end{tabular}
\caption{Russia's censorship of llama.meta.com and coursehero.com. The statistics were taken from a  April 15, 2024, to April 29, 2024. $\checkmark$ represents correct responses, $\times$ - incorrect responses (blocks), and $\sim$ for alternating between correct and incorrect responses. R represents the number of queried resovlers.}
\label{tab:russia_20485_3267_blocks}
\end{table}

%

We found different ASNs have different natures of blocking. An interesting behavior demonstrated in Table~\ref{tab:russia_20485_3267_blocks} is alternating. Some ASNs, for example, TransTeleCom (AS20485), hold resolvers that do not send the same type of answer during our probe. The resolvers alternate between correct answers and incorrect answers.   In our results, we find that all alternating resolvers shift from correct to incorrect answers in each pair of adjunct responses.
This may be a result of a load-balancing process between machines in these IP addresses~\cite{shue2013resolvers}.

Interestingly, the resolvers in Table~\ref{tab:russia_20485_3267_blocks} behave differently for the two domains. For example,
TransTeleCom (AS20485) in the Llama case, holds an almost even amount of resolvers that respond with correct and incorrect responses. Also, 18.5\% of the resolvers alternate - sometimes returning the right answer, and sometimes a bad one. In the Coursehero case, most resolvers of TransTelecom do not block the domain (73\%), and 19\% block it. 8\% alternate between correct and incorrect answers. RUNNet (AS3267) has some level of blocking of Llama (20\% block, and 10\% alternate). However, most resolvers (70\%) of RUNNet return the correct answer. For Coursehero, RUNNet's resolvers block this domain. In ER-Telecom (AS43478), only 14\% of the resolvers block Llama, but all resolvers block Coursehero. 

Other ASNs show similar behavior for the two domains. BILLING SOLUTION Ltd. (AS199945), Great St. Petersburg Polytechnic University (AS39503) and Linden Investors LP (AS16407) tend to block both Llama and Coursehero with high percentage. This is done both by full block or alternation. 

Also, for all the resolvers of TransTeleCom (AS20485) blocking is not done directly. The CNAME that is received from the authoritative resolver of llama.meta.com is star.c10r.facebook.com, this CNAME is mapped to the local CNAME fz139.ttk.ru, which it mapped to a blocking web page. 
This is the only ASN we found to follow this behavior. 
We found another unique behavior in TransTeleCom. For one resolver of TransTelecom, 
multiple resource records are returned in the same query message at three different times (namely, in about 1\% of the answers of this resolver), where some of the records are correct and others are incorrect. 

Moreover, we explored the different blocking methods in each ASN, whether the resolvers use a blocking page, a non-existent domain, a sinkhole, or other blocking techniques. 
This will be discussed next.

\subsection{Blocking Methods in Russian ASNs}
\label{rus-meth}
We extend the view of the Russian blocks of Llama and Coursehero (Section~\ref{russia-blocks-sec}). In Table~\ref{tab:russia_blocks_methods}, we list the methods used by each ASN to block the two domain names: Sinkhole (S), non-routable/internal IP (I), non-exist domain (N), block page (B), and Error (E). 
We find that some ASNs are consistent in their method (AS3267 and AS39503 - sinkhole, AS33908 with a blocking page stating the access is limited due to a Russian Federal law from 2006\footnote{https://eais.rkn.gov.ru/docs.eng/149.doc}). Others use different methods for different resolvers (e.g., AS20485 and AS8641), with block page, internal IP, errors, nonexist domains, etc.
\begin{table}[h]
\centering
\begin{tabular}{|c|c|c|}
\hline
\textbf{Domains$\Rightarrow$} & {\textbf{llama.meta.com}} & {\textbf{coursehero.com}} \\
\hline
 \textbf{ASN$\Downarrow$}& &\\
\hline
\textbf{AS3267}    &S & S \\
\hline
\textbf{AS20485}   &I/N/B/E & I/N/B/E \\
\hline
\textbf{AS8641}    &N/B&N/B \\
\hline
\textbf{AS41446} &I& - \\
\hline
\textbf{AS16047}   &S/B&S/B \\
\hline
\textbf{AS33908}   &B&B \\
\hline
\textbf{AS43478}   &S/B&S/B \\
\hline
\textbf{AS199945}  &S/B&S/B \\
\hline
\textbf{AS39503}  &S&S  \\
\hline
\end{tabular}
\caption{Blocking 
 methods of llama.meta.com and coursehero.com in popular ASNs in Russia. }
\label{tab:russia_blocks_methods}
\end{table}

In summary, in Russia, we found two domains to be partially censored - llama.meta.com and Couresehero.com. The nature of blocking of these domains is different between different ASNs, and also between resolvers of the same ASN. This shows that the inconsistency in Russia continues in the block methodology.

\section{Comparison with Other Platforms}
\label{others}
We explored results from other platforms at the same times as our measurements. The included platforms are OONI~\cite{filasto2012ooni,basso2021measuring}, Censored Planet (CP)~\cite{sundara2020censored,raman2023advancing}, and GFWatch~\cite{hoang2021great}. These platforms are not focused on Generative AIs, so there is limited data to compare with.

We manually examined the domains in Table~\ref{tab:tab-cat1} for each platform data.
OONI provided data on ChatGPT and Bard. We showed above China's censorship of ChatGPT using DNS. OONI also found DNS-based censorship for the top Chinese ASNs: China Telecom (AS4134), China Unicom (AS4837),  and China Mobile (AS9808). OONI, like us, found no DNS censorship in China for Bard, but they found censorship using TCP/IP blocking. Like us, they found
no censorship for both domains in Russia.

The only domains of this study found in CP probes are ChatGPT, Bard and Openai.com. For Russia, CP data showed 0\%-6\% of DNS censorship, but mostly no censorship was found.
For China, CP probes found DNS censorship of ChatGPT with more than 98\% for top ASNs (AS4134, AS4847, AS9808). For OpenAI and Bard, DNS censorship was not found. As with OONI, the censorship was noticeable in other protocols, i.e., HTTP/HTTPS. 

GFWatch, analyzes GFW's censorship. It included several domains from our list\footnote{ChatGPT and its parent domain OpenAI, beta.character.ai,dreamstudio.ai, civitai.com, starryai.com, openart.ai, poe.com, sdk.vercel.ai, chat.lmsys.org,
huggingface.co, coursehero.com, writesonic.com, deepai.org, novelai.net.}. 
These domains were reported as censored but without explicit timestamps.

We did not include in this comparison two platforms: ICLab~\cite{ICLab:SP20} and GFWeb~\cite{298156}. ICLab stopped measuring data after 2021, a time in which no Generative AI was accessible to the public. Thus, they did not add Generative AI domains to their measurements. 
GFWeb was introduced recently, and the data is still not accessible to compare.

\section{Conclusions and Future Work}
\label{conclusion}
In this work, we presented the first measurement of Generative AI censorship via DNS. We evaluated the accessibility of various Generative AI domains worldwide and found only two large-scale censorships: in China and Russia. China blocks 14 Generative AI domains, nationwide. We also show China's inconsistent behavior: e.g., blocking ChatGPT but not Claude. In Russia, censorship is limited to 2 domains Llama and CourseHero, and limited to a few ASNs. 

We are aware that our work holds some limitations. First, we only explore domains using DNS resolution, thus we can not explore inner paths within domains. This drawback is a consequence of using the system of~\BG, which is limited to such domains. It is also possible that the resolvers in China and Russia are blocking the specific cloud providers we are using.

It is quite possible that some of the resolvers we queried are non-official, namely they reside in the IP space of a provider but managed by someone else, e.g., in a VPS.  However, since in most cases, the behavior we see is seen by tens, hundreds, and thousands of resolvers this effect can not be significant.  We selected not to use the resolver importance rating supplied to us since we could not expose the ranking process.

Our future work includes extending the probe to a year-long probing, to capture trends and dynamics of censorships, especially of new domains. We would like to follow new Generative AI domains from the first time they are introduced, to capture the first time they are censored.

\bibliographystyle{IEEEtran}

\appendix
\section{Ethical Considerations}
\label{ethics}
In conducting this research, we carefully considered ethics related to the probing of DNS resolvers worldwide. To perform the DNS queries, we employed a tool provided by \BG, utilizing proxies located across the globe to facilitate large-scale query campaigns. The vast majority of these queries were executed through proxies residing within three major cloud providers, situated in various locations across Europe and the USA. Only a small fraction (approximately 0.5\%) of queries were conducted via residential proxies. We ensured that no queries were made from residential areas in sensitive locations such as China or Russia, where users may be potentially exposed to consequences employed by their regime due to directly probing forbidden DNS infrastructure.

\end{document}